\documentclass[11pt]{article}

\usepackage{color,graphicx}
\usepackage{young}
\usepackage[vcentermath]{youngtab}
\usepackage{amsmath,amssymb,graphicx}
\usepackage{hyperref}
\definecolor{darkred}{rgb}{0.65,0.15,0}
\hypersetup{pdfborder={0 0 0},colorlinks=true,urlcolor=darkred,citecolor=blue,linkcolor=darkred,linktocpage=true}

\usepackage{cite}
\usepackage{amsmath}
\usepackage{amsfonts}
\usepackage{amssymb}
\usepackage{graphicx}%
\usepackage{amsthm}
\usepackage{mathrsfs}
\usepackage[T1]{fontenc}
\usepackage{enumerate}
\usepackage{color}
\usepackage{hyperref}
\hypersetup{
   colorlinks   =  true,
    citecolor    = red,
     urlcolor	=magenta,
}
\def\4diml{four-dimensional}

\def\-1{^{-1}}

\newcommand{\G}{\mathscr{G}}

\makeatletter

\@addtoreset{equation}{section}
\makeatother

\begin{document}

\thispagestyle{empty}

\vspace{5mm}

\begin{center}
{\LARGE \bf Yang-Baxter deformations of WZW model on \\[2mm] the Heisenberg Lie group}

\vspace{14mm}
\normalsize
{\large   Ali Eghbali\footnote{eghbali978@gmail.com}, Tayebe Parvizi\footnote{t.parvizi@azaruniv.ac.ir},
Adel Rezaei-Aghdam\footnote{Corresponding author: rezaei-a@azaruniv.ac.ir}}

\vspace{4mm}
{\small {\em Department of Physics, Faculty of Basic Sciences,\\
Azarbaijan Shahid Madani University, 53714-161, Tabriz, Iran}}\\

%\hrule

\vspace{10mm}

\begin{tabular}{p{12cm}}
{\small
The Yang-Baxter (YB) deformations of Wess-Zumino-Witten (WZW) model on the Heisenberg Lie group ($H_4$) are examined.
We proceed to obtain the nonequivalent solutions of (modified) classical Yang-Baxter equation ((m)CYBE) for the $ h_4$ Lie algebra by using
its corresponding automorphism transformation.
Then we show that YB deformations of $H_4$ WZW model
are split into ten nonequivalent backgrounds including metric and $B$-field
such that some of the metrics of these backgrounds can be transformed to the metric of $H_{4}$ WZW model while the antisymmetric $B$-fields are changed.
The rest of the deformed metrics have a different isometric group structure than the $H_{4}$ WZW model metric.
As an interesting result, it is shown that all new integrable backgrounds of the YB deformed $ H_4$ WZW model
are conformally invariant up to two-loop order. In this way, we obtain the general form of the dilaton fields satisfying the
vanishing beta-function equations of the corresponding  $\sigma$-models.
}
\end{tabular}
\vspace{6mm}
%\hrule
\end{center}

\newpage
\setcounter{page}{1}

\tableofcontents

 \vspace{5mm}
% \hrule
 \vspace{5mm}

\section{\label{Sec.II} Introduction}

The study of integrable two dimensional $\sigma$-models and their deformations
have always remarkable attentions of people from early times of their presentation \cite{chiral,linear}.
Integrable deformations of $SU(2)$ principal chiral model firstly presented in
\cite{family,Relativistically,representation }.
Then, YB (or $\eta $) deformation of chiral model
was introduced by Klimcik \cite{klimcik2003yang,klimvcik2009integrability,klimvcik2014}
as the generalization of \cite{Relativistically,representation} while the model proposed in\cite{family}
was generalized as $\lambda$-deformation in \cite{interpolations}.
The relation between these integrable deformations was studied in \cite{sfetsos2015generalised,klimcik2015eta}.
The YB integrable deformations \cite{klimcik2003yang} are based on $R$-operators satisfying
the mCYBE and the generalization to models with $R$-operators satisfying the CYBE (homogeneous YB deformations)
was also studied in \cite{matsumoto2015yang}. The application of these integrable deformation to
string theory specially the AdS$_5\times$S$^5$ string model has presented  in \cite{delduc2014integrable,Jordanian,supercoset}
(see also \cite{HidekiKyono,Arutyunov1,Hoare1}).
For homogeneous YB deformed models
it has been shown that \cite{supergeometry}
there is no Weyl anomaly if the $R$-operators are unimodular (see also \cite{Two-loop} up to two-loop, and \cite{Hronek}).
In \cite{Sheikh-Jabbari1}, the relationship between unimodularity condition on $R$-matrices with the divergence-free of the
noncommutative parameter $\Theta$ of the dual noncommutative gauge theory has been discussed; moreover, it has been shown that \cite{Sheikh-Jabbari2}
the equations of motion of the generalized supergravity reproduce the CYBE in such a way that $\Theta$ is the most general $r$-matrix solution
built from antisymmetric products of Killing vectors.
The $r$-matrices may be divided into Abelian and non-Abelian,
and it has been proved that Abelian $r$-matrices correspond to T-duality shift T-duality transformations \cite{Tongeren},
thus ensuring that the corresponding YB deformation is a supergravity solution.
In the case of  non-Abelian $r$-matrices,
the unimodularity condition on the $r$-matrix \cite{supergeometry}  distinguishes valid
supergravity backgrounds \cite{callan1985strings} from the generalized supergravity solutions \cite{invariance,generalized}.
The Weyl invariance of bosonic string theories
on generalized supergravity backgrounds was shown at one-loop order by
constructing a local counterterm \cite{Sakamoto1,Sakamoto2}.

The generalization to YB $\sigma$-models with WZW term has also
carried out in \cite{delduc2015integrable,kyono2016yang,Klimcik2017,Quantum,2020}.
In most of the works, the models have been constructed on semisimple or compact Lie groups.
In Ref. \cite{kyono2016yang}, the YB models on the Nappi-Witten group was constructed.
There, it has been shown that the Nappi-Witten model is the unique conformal theory within the class of the YB
deformations preserving the conformal invariance.
Lately, YB deformation of the Nappi-Witten background based on the mCYBE has been used in order to find a one-parameter family of supergravity
solutions which contains the Nappi-Witten background and the flat Minkowski space \cite{Sakatani}.
Here we particularly focus on the YB $\sigma$-models with WZW term on the $ H_4$ Lie group
obtaining from $R$-operators satisfying the (m)CYBE.
We show that YB deformations of $H_4$ WZW model
are split into ten nonequivalent backgrounds including metric and $B$-field such that
some of the metrics of these backgrounds can be transformed to the metric of  $H_{4}$ WZW model while the antisymmetric $B$-fields are changed.

The plan of the paper is as follows:
In order to present the notations,  we review in general the YB deformations of chiral and WZW models in Sec. \ref{Sec.II}.
In Sec. \ref{Sec.III}, after a review of the construction of WZW model based on the $ H_4$ Lie group
\cite{kehagias1994exact,eghbali2015poisson},
by using the automorphism group of the $h_4$ Lie algebra
we obtain the solutions of the (m)CYBE,
i.e. corresponding nonequivalent classical r-matrices.
We prove that in general the equivalent classical r-matrices
(r-matrices related by automorphism group) lead to equivalent models.
After then, we classify all backgrounds of YB deformed WZW model on $H_4$ in subSec. \ref{subSec.III.3}.
The use of the convenient coordinate transformations (similar to YB deformed WZW model on the Nappi-Witten group \cite{kyono2016yang})
in order to transform the metrics of some deformed backgrounds to the metric of $H_4$ WZW model is given at
the end of Sec. \ref{Sec.III}.
The one-loop conformal invariance of the deformed models is investigated in subSec. \ref{subSec.IV.1} in such a way that the
corresponding dilaton fields are found. In subSec. \ref{subSec.IV.2}, we immediately check the conformal invariance of the models up to two-loop order and
conclude that two-loop beta-function equations are satisfied with the same previous dilaton fields.
Some concluding remarks are given in the last section.
We tabulate the nonzero components of tensors $ H_{\mu\nu\rho}, (H^{2})_{\mu\nu}, R_{\mu \nu}$ and Riemann tensor field
related to the backgrounds of YB deformed $H_4$ WZW model in Appendix \ref{app.A}.
Finally, in Appendix \ref{app.B}, by following our present method
we classify all nonequivalent classical r-matrices and corresponding YB deformed WZW models based on
the Nappi-Witten group \cite{kyono2016yang}; moreover, we show that all deformed backgrounds are conformally invariant up to two-loop order.

%%%%%%%%%%%%%%%%%%%%%%%%%%%%%%%%%%%%%%%%%%%%%%%%%%%%%%%%%%%%%%%%%%%%%%%%%%%%%%%%%

\section{\label{Sec.II} A review of YB $\sigma$-model and YB deformations of WZW model}

Before proceeding to review the YB deformations of WZW model, let us introduce
the YB deformation of the principal chiral model on the Lie groups.

\subsection{\label{subSec.II.1}YB $\sigma$-model}

In order to make the paper somewhat self-contained, let us first start with the YB deformation
of the principal chiral model on a Lie group $G$ (with Lie algebra ${\G}$), giving \cite{klimcik2003yang}
\begin{eqnarray}\label{2.1}
S=-\frac{1}{2}\int_{_\Sigma} d^{2}\sigma Tr\Big[(g^{-1} \partial_{-}g)\frac{1}{1-\eta R} (g^{-1} \partial_{+}g)\Big],
\end{eqnarray}
where $\partial_{\pm}=\partial_{\tau} \pm \partial_{\sigma}$ are the derivatives with respect to the standard lightcone
variables $\sigma^{\pm} = (\tau \pm \sigma)/2$ on the worldsheet $\Sigma$, and
$g^{-1} \partial_{\pm}g$ are components of the left-invariant Maurer-Cartan one-forms
which are defined by means of an element $g: \Sigma \rightarrow G$ in the following formula
\begin{eqnarray}\label{2.2}
g^{-1} \partial_{_{\pm}} g \equiv {L_{_{\pm}}} ={L_{_{\pm}}^{i}}~T_{{_i}},
\end{eqnarray}
in which $T_{{_i}}, i=1,...,dim \hspace{0.4mm}G$ are the bases of Lie superalgebra ${\G}$.
In Eq. \eqref{2.1}, $\eta$ is a real parameter by which deformation is measured.
If one puts $\eta=0$, the action reduces to the principal chiral model \cite{chiral,linear}.
In addition, the linear operator\footnote{One can associate the $R$-operator to a classical r-matrix \cite{matsumoto2015yang}.}{$R: {\G} \rightarrow {\G}$ is the solution of equation \cite{matsumoto2015yang}
\begin{eqnarray}\label{2.3}
[R(X),R(Y)]-R\big([R(X),Y]+[X,R(Y)]\big)=\omega [X,Y],
\end{eqnarray}
for all $X, Y \in {\G}$. Here $\omega$ is constant parameter which can be normalized by rescaling $R$.
When $\omega=0 $, the equation \eqref{2.3} is called the CYBE. This equation can be generalized to
the mCYBE with $\omega=\pm1$.
The skew-symmetric condition of operator $R$ is written as
\begin{eqnarray}\label{2.4}
Tr(R(X) Y)+Tr(X R(Y))=0.
\end{eqnarray}
The integrability of the model \eqref{2.1} is an important property of the model
such that the corresponding Lax pair is given by\footnote{Note that
the Lax pair in \eqref{2.5} is the one for $\omega=0$. One can find a general form of the Lax pair for an arbitrary $\omega$
rather than \eqref{2.5}, giving \cite{matsumoto2015yang}
\begin{eqnarray*}
{{\cal L}}_{\pm}(\lambda)=\frac{1}{1\pm\lambda}\Big(1 \mp \frac{\lambda\eta(\eta \omega \pm R)}{1\pm\eta R}\Big){L_{_{\pm}}}.
\end{eqnarray*}
} \cite{matsumoto2015yang}
\begin{eqnarray}\label{2.5}			
{{\cal L}}_{\pm}(\lambda)=\frac{1}{1\pm\lambda}\big(1 -\frac{\lambda\eta R}{1\pm\eta R}\big) {L_{_{\pm}}},
\end{eqnarray}
where $\lambda $ is a spectral parameter.

\subsection{\label{subSec.II.2} YB deformation of WZW model}

In this subsection we shall consider the YB deformation of the WZW model \cite{delduc2015integrable}.
The corresponding action consists of standard principal chiral model and WZW term based
on a Lie group $G$, giving \cite{{delduc2015integrable},{kyono2016yang}}
\begin{eqnarray}\label{2.6}			
S^{^{YB }}_{_{WZW}}(g)=\frac{1}{2}\int_{_\Sigma}d^{2}\sigma ~\Omega_{ij} L_{-}^{i} J_{+}^{j} +\frac{\kappa}{2}\
\int_{_{B_{3}}}d^{3}\sigma  ~\Omega_{kl}~ f_{ij}^{~l} ~ L_{\xi}^{i} L_{+}^{j} L_{-}^{k},
\end{eqnarray}
in which $\kappa$ is a constant parameter, {\small $B$} is a three-manifold bounded by worldsheet $\Sigma$, and
$\Omega_{kl}$ defined by $\Omega_{kl} =<T_k , T_l>$ is a non-degenerate ad-invariant symmetric bilinear form on Lie algebra ${\G}$
with structure constants $f_{ij}^{~~k}$ which satisfies the following relation \cite{nappi1993wess}
\begin{eqnarray}\label{2.7}
f_{ij}^{\;\;l} \;\Omega_{lk}+ f_{ik}^{\;\;l} \;\Omega_{lj}\;=\;0.
\end{eqnarray}
Here the deformed currents $J_{\pm}$ are defined in the following way
\begin{eqnarray}\label{2.8}
J_{\pm} = (1+\omega \eta^{2})\frac{1 \pm \tilde{A} R}{1-\eta^{2}R^{2}} L_{\pm},
\end{eqnarray}
where $\eta$ and $\tilde{A}$  measure a deformation of WZW model.
One can see that when $\eta = \tilde{A}=0$ and $ k=0$($k=1$) we recover the action of the
principal chiral model(undeformed WZW model) \cite{delduc2015integrable},
and for $ \tilde{A}=\pm \eta, k=0 $ one recovers the YB deformation of chiral model \cite{chiral,linear}.
In general, the constant parameter $\omega$ classifies integrable deformations so that one may consider $ \omega  = 0, \pm 1$ \cite{chiral,linear}.
In \cite{delduc2015integrable}, it was shown that in general the model \eqref{2.6} is integrable.
This model was then considered for the Nappi-witten group \cite{kyono2016yang}. In the next section,
we will consider the model \eqref{2.6} for the $H_{4}$ Lie group.

%%%%%%%%%%%%%%%%%%%%%%%%%%%%%%%%%%%%%%%%%%%%%%%%%%%%%%%%%%%%%%%%%%%%%%%%%%%%%%%%%%%%%%%%%%

\section{\label{Sec.III} YB deformations of WZW model based on the $H_{4}$ Lie group and their classification}

In this section, we shall solve the mCYBE to obtain the classical r-matrices of the $h_4$ Lie algebra. Since
our goal is the classification of all nonequivalent r-matrices,
we prove a Proposition. This Proposition states that two r-matrices $r$ and $r'$ equivalent if one can be obtained from the other by means of
a change of basis which is an automorphism $A$ of Lie algebra $\G$.
We then calculate all linear $R$-operators corresponding to nonequivalent r-matrices in order to construct
the YB deformations of the $H_{4}$ WZW model.
Finally, by performing convenient coordinate transformations on {\it some of the deformed backgrounds}
we show the invariance of the $H_{4}$ WZW model metric under arbitrary YB deformations, up to antisymmetric $B$-fields.
This means that the effect coming from the deformations is reflected only as the coefficient of $B$-field.

\subsection{\label{subSec.III.1} WZW model based on the $H_{4}$ Lie group}

In this subsection we shall consider the WZW  model on the $H_{4}$ Lie group \cite{kehagias1994exact,eghbali2015poisson}.
Before proceeding to construct model, let us first introduce the $h_4$ Lie algebra of $H_4$.
The Lie algebra $h_4$ is defined by the set of generators $(T_{1}, T_{2}, T_{3}, T_{4})$ with the following nonzero Lie brackets
\begin{eqnarray}\label{3.1}
[T_{1} , T_{2}]~=~T_{2},~~~~~[T_{1} ,T_{3}]~=~-T_{3},~~~~~[T_{3} , T_{2}]~=~T_{4}.
\end{eqnarray}
The action of ungauged and undeformed WZW model on a Lie group $G$ is given by
\begin{eqnarray}\label{3.2}
S_{_{WZW}}(g) &=&  \frac{1}{2} \int_{_\Sigma} d\sigma^+ d\sigma^-\;{\Omega}_{ij}
			L^{i}_{+}\;
			L^{j}_{-}
			+\frac{1}{12} \int_{_B} d^3 \sigma~
			\varepsilon^{ \gamma \alpha \beta}~{\Omega}_{ik} \;f_{jl}^{~~k}~
			L^{i}_{_\gamma} L^{j}_{_\alpha} L^{l}_{_\beta}.
		\end{eqnarray}
Accordingly, one needs a non-degenerate bilinear form $\Omega_{ij}$ on Lie algebra ${\G}$ of $G$.
Using \eqref{3.1} and also formula \eqref{2.7}, one can get
the non-degenerate bilinear form on the $h_4$, giving \cite{eghbali2015poisson}
\begin{eqnarray}\label{3.3}
\Omega_{ij}=\left( \begin{tabular}{cccc}
                 $\rho$ & 0 & 0 & -$\lambda$ \\
                 0 & 0 & $\lambda$ & 0 \\
                 0 & $\lambda$& 0 & 0 \\
                 -$\lambda$ & 0 & 0 & 0 \\
                 \end{tabular} \right),
\end{eqnarray}
where $\rho$ and $\lambda$ are some real constants.
To construct the WZW action \eqref{3.2} on the $H_4$, we parameterize an element of the $H_4$ as
\begin{eqnarray}\label{3.4}
g=e^{v T_{_4}} ~ e^{u T_{_3}} ~e^{x T_{_1}} ~e^{y T_{_2}},
\end{eqnarray}
where $x^{\mu} =(x, y, u, v)$ stand for the coordinates of the $H_4$ group manifold.
Using \eqref{3.1} and \eqref{3.4} the corresponding left-invariant one-forms components are obtained to be \cite{eghbali2015poisson}
\begin{eqnarray}\label{3.5}
L^{\hspace{-0.5mm}1}_{\pm}=\partial_{\pm} x,~~~~~L^{\hspace{-0.5mm}2}_{\pm}=y \partial_{\pm} x + \partial_{\pm} y,~~~~~
L^{\hspace{-0.5mm}3}_{\pm}=e^{x}~\partial_{\pm} u,~~~~~L^{\hspace{-0.5mm}4}_{\pm}=ye^{x}~ \partial_{\pm} u + \partial_{\pm} v.
\end{eqnarray}
Finally, the WZW action on the $H_4$ is found to be of the form \cite{eghbali2015poisson}
\begin{eqnarray}\label{3.6}
S_{_{WZW}}(g)&=&\frac{1}{2}\int d\sigma^+d\sigma^-\Big[\rho~ \partial_+x \partial_-x-\partial_+x \partial_-v -\partial_+v \partial_-x
			+e^x\big(\partial_+y \partial_-u+\partial_+u \partial_-y\nonumber\\
			&&~~~~~~~~~~~~~~~~~~~~~+y\partial_+u \partial_-x- y\partial_+x \partial_-u\big)\Big].
\end{eqnarray}
Here we have set $\lambda=1$.
Identifying the action \eqref{3.6} with the $\sigma$-model
of the form\footnote{$h_{\alpha \beta}$  and  $\epsilon^{\alpha \beta}$ are
the induced metric and antisymmetric tensor on the worldsheet, respectively, such that
$h= \det {h_{\alpha \beta}}$ and the indices $\alpha, \beta$ run over $(\tau, \sigma)$.
The dimensionful coupling constant $\alpha'$ turns out to be the inverse string tension.}
\begin{eqnarray}\label{3.7}
S=\frac{1}{4 \pi \alpha'}\int_{_{\Sigma}}\!d\tau  d\sigma \sqrt{-h} \big(h^{\alpha \beta}G_{_{\mu\nu}}
			+\epsilon^{\alpha \beta} B_{_{\mu\nu}}\big)\partial_{\alpha}x{^{^\mu}}
			\partial_{\beta}x^{^{\nu}} ,
\end{eqnarray}
we can read off the spacetime metric $G_{_{\mu\nu}}$ and the antisymmetric $B$-field. They are then given by the following relations
\begin{eqnarray}
ds^2 &=&\rho~d x^{2}-2~ dx~ dv + 2 e^x~ dy~ du,\label{3.8}\\
B&=&-ye^x~ dx \wedge du.\label{3.8.1}
\end{eqnarray}
The metric \eqref{3.8} has an isometry group, where the generators of the corresponding Lie algebra
can be expressed in terms of the Killing vectors $K_i$ of the target space geometry.
Therefore it is crucial for our further considerations to obtain the Lie algebra of Killing vectors of \eqref{3.8}.
This metric admits a seven-dimensional Lie algebra of Killing vectors, which can be generated by
\begin{eqnarray}\label{3.8.2.1}
K_1 &=&- \frac{\partial}{\partial x}+ u \frac{\partial}{\partial u}- \rho  \frac{\partial}{\partial v},~~~~~~~~~~~~~~~~
K_2=e^{-x} \frac{\partial}{\partial y} -u \frac{\partial}{\partial v},\nonumber\\
K_3&=& \frac{\partial}{\partial y},~~~~~~~~~~~~~~~~~~~~~~~~~~~~~~~~~~~~~K_4=y \frac{\partial}{\partial y} -u \frac{\partial}{\partial u},\nonumber\\
K_5&=& e^{-x} \frac{\partial}{\partial u} -y \frac{\partial}{\partial v},~~~~~~~~~~~~~~~~~~~~~~~K_6=\frac{\partial}{\partial u},\nonumber\\
K_7&=& -\frac{\partial}{\partial v}.
\end{eqnarray}
One can easily check that the Lie algebra spanned by these vectors is
\begin{eqnarray}\label{3.8.2}
{[K_1 , K_2]} &=&K_2,~~~~~~~[K_1 , K_6] = -K_6,~~~~~~~[K_2 , K_4]=K_2,~~~~~~~~[K_2 , K_6] = -K_7,\nonumber\\
{[K_3 , K_4]} &=&K_3,~~~~~~~[K_3 , K_5] = K_7,~~~~~~~~~[K_4 , K_5] = K_5,~~~~~~~~~[K_4 , K_6] = K_6,
\end{eqnarray}
with the center $K_7$. The generator $K_4$ can be interpreted as dilation in $y, u$. As it is seen, the $h_4$ Lie algebra,
e.g. generated by $(K_1 , K_2, K_6 , K_7)$,  is a subalgebra of \eqref{3.8.2}.
%%%%%%%%%%%%%%%%%%%%%%%%%%%%%%%%%%%%%%%%%%%%%%%%%%%%%%%%%%%%%%%%%%%%%%%
\subsection{\label{subSec.III.2} Classical r-matrices for $h_{4}$ Lie algebra}

According to the formulas \eqref{2.6} and \eqref{2.8}, to obtain the YB deformations of the
$H_4$ WZW model one needs the linear operators $R$ associated to classical r-matrices of the $h_4$ Lie algebra.
Before proceeding to this, let us consider the general form of classical r-matrix of
a given Lie algebra $\G$ with the basis $\{T_i\}$ \cite{rezaei2005classification}
\begin{eqnarray}\label{3.9}
r=\frac{1}{2}r^{ij}(T_i \otimes T_j-T_j \otimes T_i),
\end{eqnarray}
where $r^{ij}$ is an antisymmetric matrix. One may associate a linear operator $R$ to a r-matrix
that satisfies the mCYBE \eqref{2.3}. This operator can be defined in the following way \cite{kyono2016yang}
\begin{eqnarray}\label{3.10}
R(T_{k})=<r, (1\otimes T_{k})> = r^{ij} \Omega_{jk} ~ T_{i}.
\end{eqnarray}
Based on this, the action of $R$ on any element $X=x^k T_{k} \in \G$ is written as
\begin{eqnarray}\label{3.11}
R(X)= x^k  R(T_{k})= r^{ij} \Omega_{jk} x^k ~ T_{i}.
\end{eqnarray}
Considering
\begin{eqnarray}\label{3.12}
R(T_{k}) = {R_k}^{i} ~ T_{i},
\end{eqnarray}
and then comparing \eqref{3.10} and \eqref{3.12}, one gets
\begin{eqnarray}\label{3.13}
{R_k}^{i} =  r^{ij} \Omega_{jk}.
\end{eqnarray}
Now, making use of formulas \eqref{2.7} and \eqref{3.11} and after some algebraic calculations,
one can write Eq. \eqref{2.3} in the following form \cite{kyono2016yang}
\begin{eqnarray}\label{3.14}
{f_{lm}}^k r^{li} r^{mj}+{f_{lm}}^i r^{lj} r^{mk}+{f_{lm}}^j r^{lk} r^{mi}
-\omega {f_{lm}}^k \Omega^{li} \Omega^{mj}=0.
\end{eqnarray}
This equation can be used in order to calculate the r-matrices for a given Lie algebra $\G$.
But, for obtaining the nonequivalent r-matrices one must use the automorphism group of Lie algebra $\G$.
The action of the automorphism $A$ on ${\G}$ is given by the following transformation
\begin{eqnarray}\label{3.15}
T'_i = {A}(T_i) =  {A}_i^{~j}~ T_j,
\end{eqnarray}
where $T'_i$ are the changed basis by the automorphism $A$.
Since the automorphism preserves the structure constants, the basis $T'_i$ must obey the same commutation relations as
$T_i$, i.e.,
\begin{eqnarray}\label{3.16}
[T'_i ,  T'_j] = {f_{_{ij}}}^k ~T'_k.
\end{eqnarray}
Inserting the transformation \eqref{3.15}  into \eqref{3.16}  we find that the elements of
automorphism group $A$ satisfy the following relation
\begin{eqnarray}\label{3.16.1}
{A}_i^{~m}~ {f_{_{mn}}}^k~ {A}_j^{~n} = {f_{_{ij}}}^l~ A_l^{\;\;k}.
\end{eqnarray}
In order to calculate the elements ${A}_i^{~j}$ of Lie algebra $\G$ it would be helpful to further write
the matrix form of \eqref{3.16.1}, giving\footnote{Here ``t'' denotes transposition.} \cite{ER1}
\begin{eqnarray}\label{3.17}
A~ {\cal Y}^k A^{t} = {\cal Y}^l A_l^{\;\;k},
\end{eqnarray}
where $(\mathcal{Y}^{k})_{ij}=- {f_{ij}}^k$ are the adjoint representations of $\G$.
It is also useful to obtain matrix form of Eq. \eqref{3.14} by using the adjoint representations $(\mathcal{Y}^{k})_{ij}=- {f_{ij}}^k$
and $({\cal X}_{i})_{j}^{~k}=-{f_{ij}}^k$.
It is then read
\begin{eqnarray}\label{3.18}
r \mathcal{Y}^{k} r+ r (\mathcal{X}_{l} r^{lk})- (r^{kl} \mathcal{X}_{l}^{t}) r=-\omega(\Omega^{-1} \mathcal{Y}^{k} \Omega^{-1}).
\end{eqnarray}
In order to determine the nonequivalent r-matrices for a given Lie algebra $\G$ we give Proposition 3.1.\\\\
{\bf Proposition 3.1.} {\it Let $r$ and $r'$ be two r-matrices as solutions of the (m)CYBE \eqref{3.14}.
If there exists an automorphism $A$ of $\G$ such that
\begin{eqnarray}\label{3.19}
r = A^t~ r^\prime~A,
\end{eqnarray}
then the matrices $r$ and $r'$ of Lie algebra $\G$ are equivalent. }\\
{\bf Proof.}~Let $\{T_i\}$ and $\{T'_i\}$ be the bases of $\G$ such that $T'_i = {A}_i^{~j}~ T_j$ in which ${A}_i^{~j}$
is an element of automorphism group  $Aut (\G)$.
Since an automorphism $A$ of $\G$ preserves the structure constants, one may use \eqref{2.7} to conclude that $\Omega_{ij} =<T'_i , T'_j>$.
Then, using \eqref{3.15} it is simply shown that
\begin{eqnarray}\label{3.20}
{A}_i^{~k}~\Omega_{kj} = \Omega_{il}~ (A^{-1})_j^{~~l}.
\end{eqnarray}
On the one hand, according to \eqref{3.12} for the changed basis we find $R(T'_i)={R'}_i^{~j} ~T'_j={R'}_i^{~j}~{A}_j^{~k} ~T_k$.
In addition, one can write $R(T'_i)= {A}_i^{~l}~{R}_l^{~k}~ T_k$. Putting these relations together, one obtain that
\begin{eqnarray}\label{3.21}
{R'}_i^{~j} = {A}_i^{~k}~{R}_k^{~l}~(A^{-1})_l^{~j},
\end{eqnarray}
on the other hand, one may use \eqref{3.13} and \eqref{3.20} to write \eqref{3.21} as
\begin{eqnarray}\label{3.22}
{r'}^{jn}~\Omega_{ni} &=& {A}_i^{~k}~ \Omega_{kp}~{r}^{lp}~(A^{-1})_l^{~j}\nonumber\\
 &=&\Omega_{in}~(A^{-1})_p^{~n}~{r}^{lp}~(A^{-1})_l^{~j}.
\end{eqnarray}
Multiplying both sides of the above equation in $\Omega^{im}$, we finalize that
\begin{eqnarray}\label{3.23}
{r'}^{jm}= (A^{-1})_l^{~j}~{r}^{lp}~(A^{-1})_p^{~m},
\end{eqnarray}
and this is nothing but \eqref{3.19}. One can show that the $r'$ satisfies the (m)CYBE \eqref{3.18} if the $r$ be a solution of \eqref{3.18}.
We note that \eqref{3.19} is an equivalence relation.\\

In the following, we shall solve the (m)CYBE \eqref{3.14}  (or equivalently \eqref{3.18}) for $h_4$ Lie algebra to obtain the corresponding r-matrices.
In this respect, we consider two r-matrices $r$ and $r'$ equivalent if one can be obtained from the other by means of
a change of basis which is an automorphism $A$ of Lie algebra $\G$.
Indeed, the solutions that relate to each other through Eq. \eqref{3.19} are equivalent.
In fact, one can use \eqref{3.19} to obtain all nonequivalent r-matrices. Before proceeding further, let us calculate
the automorphism group of the particular Lie algebra $h_4$. Using the structure constants given by
\eqref{3.1} and then applying \eqref{3.17} the automorphism $A$ can be easily obtained.
The result is given by the following statement.\\\\
{{\bf Proposition 3.2.} {\it The automorphism groups of the $h_4$ Lie algebra
are expressed as matrices in basis $(T_1, \cdots, T_4)$ as } \cite{{christodoulakis2002automorphisms},{rezaei2010complex}}
{\small \begin{eqnarray}\label{3.24}
Aut(h_4)=\left\{ A_1=\left( \begin{tabular}{cccc}
                 1 & c & d & e \\
                 0 & a & 0 & ad \\
                 0 & 0 & b & bc \\
                 0 & 0 & 0 & ab \\
                 \end{tabular} \right),~A_2=\left( \begin{tabular}{cccc}
                 -1 & c & d & e \\
                 0 & 0  & a & -ac \\
                 0 & b & 0 & -bd \\
                 0 & 0 & 0 & -ab \\
                 \end{tabular} \right); ~ ab \neq 0\right \}
\end{eqnarray}}
{\it for some real constants $a, b, c, d, e$.}}\\

In order to solve the (m)CYBE \eqref{3.14} for $h_4$ Lie algebra, let us assume that $r^{ij}$ has the following general form:
\begin{eqnarray}\label{3.25}
r^{ij}=\left(\begin{tabular}{cccc}
                 0 & $m_1$ & $m_2$ & $m_3$ \\
                 -$m_1$ & 0  & $m_4$ & $m_5$ \\
                 -$m_2$ & -$m_4$ & 0 & $m_6$ \\
                 -$m_3$ & -$m_5$ & -$m_6$ & 0 \\
                 \end{tabular}\right),
\end{eqnarray}
for some real constants $m_1, \cdots ,m_6$.
By substituting \eqref{3.25} into \eqref{3.14} and then by using \eqref{3.1} together with \eqref{3.3},
the general solution of \eqref{3.14} is split into three
classes such that the solutions are, in terms of the constants $\lambda, \omega$ and $m_1, \cdots ,m_6$, given by
{\small \begin{eqnarray*}
r_{_1}=\left(\begin{tabular}{cccc}
                 0 &  0 & 0 & $m_3$ \\
                 0 & 0  & $\pm\sqrt{-\frac{\omega}{\lambda^2}}$ & $m_5$ \\
                 0 & $\mp\sqrt{-\frac{\omega}{\lambda^2}}$ & 0 & $m_6$ \\
                 -$m_3$ & -$m_5$ & -$m_6$ & 0 \\
                 \end{tabular}\right), ~~r_{_{2}}=\left(\begin{tabular}{cccc}
                 0 & $m_1$  & 0 & -$\Delta_{16}$ \\
                 -$m_1$ & 0  & $\Delta_{16}$ & $m_5$ \\
                 0 & -$\Delta_{16}$  & 0 & $m_6$ \\
                 $\Delta_{16}$  &-$m_5$ &-$m_6$ & 0 \\
                 \end{tabular}\right),
\end{eqnarray*}}
{\small \begin{eqnarray}\label{3.26}
r_{_{3}}=\left(\begin{tabular}{cccc}
                 0 & 0  & $m_2$ & $\Delta_{25}$ \\
                 0 & 0  & $\Delta_{25}$ & $m_5$ \\
                 -$m_2$ & -$\Delta_{25}$  & 0 & $m_6$ \\
                -$\Delta_{25}$ &-$m_5$ &-$m_6$ & 0 \\
                 \end{tabular}\right),~~~~~~~~~~~~~~~~~~~~~~~~~~~~~~~~~~~~~~~~~~~~~~~~~~~~~~~~~~~~~
\end{eqnarray}}
where $\Delta_{16} = \sqrt{m_1 m_6 -\frac{\omega}{\lambda^2}} $  and $\Delta_{25} = \sqrt{m_2 m_5 -\frac{\omega}{\lambda^2}} $ for all
$\omega$ in $\mathbb{R}$.
Now by using the automorphisms group elements $ A \in Aut ({h_4}) $ of \eqref{3.24}
and by employing formula \eqref{3.19} of Proposition 3.1, one concludes that r-matrices
given by \eqref{3.26} are split into ten nonequivalent
classes such that the results\footnote{In Ref. \cite{Ballesteros},
all coboundary Lie bialgebras of the $h_4$ Lie algebra have been obtained and classified into three multiparametric families.
Accordingly, their corresponding r-matrices have been also found as multiparametric.
Here we have exactly found the r-matrices of $h_4$. However, our results are in agreement with those of Ref. \cite{Ballesteros}.} are summarized in Theorem 3.1. \\\\
{\bf Theorem 3.1.}~{\it Any r-matrix of the $h_4$ Lie algebra as a solution the (m)CYBE \eqref{3.14} belongs
just to one of the following ten nonequivalent classes}
{\small \begin{eqnarray*}
r_{_I}=\left( \begin{tabular}{cccc}
                 0 & 0 & 0 & 0 \\
                 0 & 0 & 0 & 1 \\
                 0 & 0 & 0 & 0 \\
                 0 & -1 & 0 & 0 \\
                 \end{tabular} \right),~r_{_{II}}=\left( \begin{tabular}{cccc}
                 0 & 0 & 0 & 0 \\
                 0 & 0 & 0 & 1 \\
                 0 & 0 & 0 & 1 \\
                 0 & -1 &-1 & 0 \\
                 \end{tabular} \right),~r_{_{III}}=\left( \begin{tabular}{cccc}
                 0 & 1 & 0 & 0 \\
                 -1& 0 & 0 & 0 \\
                 0 & 0 & 0 & 0 \\
                 0 & 0 & 0 & 0 \\
                 \end{tabular} \right),~r_{_{IV}}=\left( \begin{tabular}{cccc}
                 0 & 0 & 0 & 1 \\
                 0 & 0 & 0 & 0 \\
                 0 & 0 & 0 & 0 \\
                 -1 & 0 & 0 & 0 \\
                 \end{tabular} \right),
\end{eqnarray*}}
{\small \begin{eqnarray*}
r_{_{V}}=\left( \begin{tabular}{cccc}
                 0 & 0 & 0 & -$1$ \\
                 0 & 0 & $1$ & 1 \\
                 0 & -$1$ & 0 & 0 \\
                 $1$ & -1 & 0 & 0 \\
                 \end{tabular} \right),r_{_{VI}}=\left( \begin{tabular}{cccc}
                 0 & 1 & 0 & -1 \\
                 -1 & 0 & 1 & 0 \\
                 0 & -1 & 0 & 0 \\
                 1 & 0 & 0 & 0 \\
                 \end{tabular} \right),r_{_{VII}}=\left( \begin{tabular}{cccc}
                 0 & 1 & 0 & 0 \\
                 -1 & 0 & 0 & 0 \\
                 0 & 0 & 0 & 1 \\
                 0 & 0 & -1 & 0 \\
                 \end{tabular} \right),r_{_{VIII}}=\left( \begin{tabular}{cccc}
                 0 & 0 & 0 & 0 \\
                 0 & 0 & 1 & 0 \\
                 0 & -1 & 0 & 0 \\
                 0 & 0 & 0 & 0 \\
                 \end{tabular} \right),~~
\end{eqnarray*}}
{\small \begin{eqnarray*}
r_{_{IX}}=\left( \begin{tabular}{cccc}
                 0 & 0 & 0 & $q^2$ \\
                 0 & 0 & 1 & 0 \\
                 0 & -1 & 0 & 0 \\
                 -$q^2$ & 0 & 0 & 0 \\
                 \end{tabular} \right),~~r_{_{X}}=\left( \begin{tabular}{cccc}
                 0 & 0 & 0 & 1 \\
                 0 & 0 & 1 & 0 \\
                 0 & -1
                  & 0 & 0 \\
                 -1 & 0 & 0 & 0 \\
                 \end{tabular} \right),~~~~~~~~~~~~~~~~~~~~~~~~~~~~~~~~~~~~
                 ~~~~~~~~~~~~~~~~~~~~~~~~~~
\end{eqnarray*}}
{\it where $q^2 \neq 0, 1$. }

It should be noted that:
\begin{itemize}

\item Both the solutions $r_{_{I}}$ and $r_{_{II}}$ can be obtained from the matrix $r_{_3}$ by putting
$ \omega=0, m_2=m_6=0, m_5=1$ and $\omega=0, m_2=0, m_5=m_6=1$, respectively; moreover, one can obtain
$r_{_{X}}$ from $r_{_3}$ by putting $ \omega=-1, \lambda=1,  m_2=m_5 = m_6=0$.
Using \eqref{3.19} we have checked that all three of the solutions $r_{_{I}}$, $r_{_{II}}$ and $r_{_{X}}$ are, under both automorphisms $A_{1}$ and $A_{2}$, nonequivalent.

\item The $r_{_{III}}$, $r_{_{VI}}$ and $r_{_{VII}}$ are just obtained from the matrix $r_{_2}$ by putting
$\omega=0, m_5= m_6=0, m_1=1$ and  $\omega=-1, \lambda=1,  m_5= m_6=0, m_1=1$, and $\omega=\lambda=1, m_5=0, m_1=m_6=1$, respectively;
moreover, the solution $r_{_{V}}$ can be obtained from $r_{_2}$ by putting $\omega=-1, \lambda=1,  m_1=m_6 =0,  m_5=1$.
We have also checked that all four of the solutions $r_{_{III}}$, $r_{_{V}}$, $r_{_{VI}}$ and $r_{_{VII}}$ are, under both automorphisms $A_{1}$ and $A_{2}$, nonequivalent.

\item All three of the solutions $r_{_{IV}}$, $r_{_{VIII}}$ and $r_{_{IX}}$ are obtained from the matrix $r_{_1}$ by putting
$\omega=0, m_5=m_6=0, m_3=1$ and  $\omega=-1, \lambda=1,  m_3=m_5= m_6=0$, and $\omega=-1, \lambda=1, m_5=m_6=0, m_3=q^2$, respectively.
One can show that these solutions are, under both automorphisms $A_{1}$ and $A_{2}$,  nonequivalent.

\end{itemize}
According to above explanations the r-matrices $r_{_{I}}, r_{_{II}}, r_{_{III}}$ and $r_{_{IV}}$ of the $h_{4}$ Lie algebra are all solutions of CYBE with $\omega =0$ while solutions of the mCYBE are  the $r_{_{V}}, r_{_{VI}}, r_{_{VII}}, r_{_{VIII}}, r_{_{IX}}$ and $r_{_{X}}$ with $\omega = \pm 1$.
Now one may use formulas \eqref{3.3}, \eqref{3.12} and \eqref{3.13}
to obtain all linear $R$-operators corresponding to the nonequivalent r-matrices.
 $R$-operators are one of the basic tools for calculating the deformed currents $J_{\pm}$
and then constructing the YB deformed WZW models. In the next subsection, we will classify all YB deformations of the $H_{4}$  WZW  model.

Before closing this subsection, it is useful to comment on the fact that the YB deformed WZW model \eqref{2.6} is, under the automorphism
transformation \eqref{3.15}, invariant. First of all, the invariance of the left invariant one-forms $L_{\alpha}$ under \eqref{3.15} requires that
\begin{eqnarray}\label{3.28}
L'^{i}_{_\alpha}= L^{j}_{_\alpha}~ (A^{-1})_j^{~i}.
\end{eqnarray}
Then, using relations \eqref{3.16.1} and \eqref{3.20} one can deduce that the second term (WZW term) of action \eqref{2.6}
is invariant with respect to the transformation \eqref{3.15}. To investigate the invariance of
the first term of \eqref{2.6}, we need to know how the currents $J_{\pm}$ change under \eqref{3.15}.
To this end, one may write down \eqref{2.8} in the following form
\begin{eqnarray}\label{3.29}
J^i_{\pm} -\eta^{2} J^k_{\pm} ~R_k^{~l}~R_l^{~i}= (1+\omega \eta^{2})\big[L^{i}_{_\pm} \pm \tilde{A} L^{k}_{_\pm} ~R_k^{~i}\big].
\end{eqnarray}
Using \eqref{3.21} and \eqref{3.28} we find that relation \eqref{3.29} does remain invariant with respect to the transformation \eqref{3.15}
if the following relation is held
\begin{eqnarray}\label{3.30}
J'^i_{\pm}= J^j_{\pm}~ (A^{-1})_j^{~i}.
\end{eqnarray}
Finally, one verifies the invariance of the first term of \eqref{2.6} under \eqref{3.15} by applying formulas \eqref{3.20}, \eqref{3.28} together with \eqref{3.30}.

%%%%%%%%%%%%%%%%%%%%%%%%%%%%%%%%%%%%%%%%%%%%%%%%%%%%%%%%%%%%%%%%%%%%%%%%%%%%%%%%%%%%%%%%%%%%%%%%%%%%%

\subsection{\label{subSec.III.3}Backgrounds for YB deformations of the $H_{4}$ WZW model}

As was mentioned earlier, by using \eqref{3.3}, \eqref{3.12} and \eqref{3.13}
we can obtain all linear $R$-operators corresponding to the nonequivalent r-matrices  of the $h_4$ Lie algebra.
Having $R$-operators, we can find the deformed currents $J_{\pm}$ from Eq. \eqref{2.8}. In this way,
one uses \eqref{2.6} to obtain YB deformations of the $H_{4}$ WZW model.
For the sake of clarity the results obtained in this subsection are summarized in Table 1;
we display the deformed backgrounds including metric and $B$-field, together with the related comments.
It should be noted that the symbol of each background, e.g. $H^{(\kappa, \eta, \tilde{A})}_{4}.III$,
indicates the YB deformed background derived by $r_{_{{III}}}$; roman numbers $I$, $II$ etc.
distinguish between several possible deformed backgrounds of the $H_{4}$ WZW model,
and the parameters $(\kappa, \eta, \tilde{A})$  indicate the deformation ones of each background.

\subsubsection{\label{subSec.III.3.1} About of the deformed backgrounds}

{\bf The backgrounds $H^{(\kappa)}_{4}.I, H^{(\kappa, \eta)}_{4}.II$ and $H^{(\kappa, \tilde{A})}_{4}.{\small X}$}.
As it is seen from Table 1, the metrics of  $H^{(\kappa)}_{4}.I$ and $H^{(\kappa,  \tilde{A})}_{4}.{\small X}$ have not, under the deformation, been
changed, i.e. in these cases, the $H_{4}$ WZW model metric remains, under the deformation, invariant while corresponding $B$-fields have been changed.
In the case of the background $H^{(\kappa, \eta)}_{4}.II$, by shifting $\rho \rightarrow \rho'=\rho -2 \eta^2$
one can easily show that this background is the same as $H^{(\kappa)}_{4}.I$. But, considering the same values of $\rho$ in both backgrounds
we are faced with a deformed metric of the $H^{(\kappa, \eta)}_{4}.II$.\\\\
{\bf The backgrounds $H^{(\kappa, \eta, \tilde{A})}_{4}.IV, H^{(\kappa, \eta, \tilde{A})}_{4}.V, H^{(\kappa, \eta)}_{4}.VIII$ and $H^{(\kappa, \eta, \tilde{A})}_{_{4,q}}.IX$}.
It is also interesting to note the fact that under some coordinate transformations one concludes that all deformed metrics
of backgrounds $H^{(\kappa, \eta, \tilde{A})}_{4}.IV, H^{(\kappa, \eta, \tilde{A})}_{4}.V, H^{(\kappa, \eta)}_{4}.VIII$ and $H^{(\kappa, \eta, \tilde{A})}_{_{4,q}}.IX$ can be turned into the same metric of the $H_{4}$ WZW model, while corresponding $B$-fields are changed. One may show that the Lie algebra of Killing vectors corresponding to metrics of these backgrounds
is isomorphic to those of \eqref{3.8}, i.e. \eqref{3.8.2}. Accordingly,
it would be interesting to try to reveal the relation between the above backgrounds and $H_{4}$ WZW model.

By performing the following coordinate transformation
\begin{eqnarray}\label{3.31}
x^{'}=\frac{1}{1-\eta^{2}} x,~~~~~y^{'}= y ~e^{\frac{-\eta^{2}}{1-\eta^{2}}  x},~~~~u^{'} =u,~~~~v^{'}=v,
\end{eqnarray}
and also by applying $\rho'=\rho (1 - \eta^2)$, we see that the metric of the background $H^{(\kappa, \eta, \tilde{A})}_{4}.IV$ turns into
the same metric of the $H_{4}$ WZW model, while $B$-field have been changed as mentioned above.
In like manner, by using the linear transformation
\begin{eqnarray}\label{3.32}
x^{'}=x,~~~~~y^{'}= y - \frac{2\eta^{2}}{1-\eta^{2}}  x,~~~~u^{'} =u,~~~~v^{'}=v,
\end{eqnarray}
and without any shift in $\rho$, one can easily show that the metric of $H^{(\kappa, \eta, \tilde{A})}_{4}.V$ is nothing but the same \eqref{3.8}.
The background $H^{(\kappa, \eta)}_{4}.VIII$ can be also simplified by performing a coordinate transformation
\begin{eqnarray}\label{3.33}
x^{'}=({1-\eta^{2}}) x,~~~~~y^{'}= y ~e^{\eta^{2}  x},~~~~u^{'} =u,~~~~v^{'}=v.
\end{eqnarray}
After performing the transformation \eqref{3.33} and using $\rho'=\rho / (1 - \eta^2)$, the resulting metric takes the same form as in \eqref{3.8}.

\begin{center}
		\scriptsize {{{\bf Table 1.}~ YB deformed backgrounds of the $H_{4}$ WZW model$^\ast$}}
		{\scriptsize
			\renewcommand{\arraystretch}{1.5}{
			\begin{tabular}{p{1.5cm}ll} \hline \hline
				Background symbol & ~~~~Backgrounds including metric and $B$-field & Comments\\ \hline
				{$H^{(\kappa)}_{4}.I$} &  ~~~~$ds^{2}=\rho dx^{2}-2dx dv+ 2e^{x} dy du,$ & \\&  ~~~~$B=\kappa y e^{x} du\wedge dx$ &$\omega=0,~\lambda=1$\\
				{$H^{(\kappa, \eta)}_{4}.II$} &  ~~~~$ds^{2}= (\rho-2\eta^{2})dx^{2}-2  dx dv+ 2 e^{x} dy du, $ & \\&   ~~~~$B=\kappa y e^{x} du\wedge dx$ & $\omega=0,~\lambda=1$ \\
	
				{$H^{(\kappa, \eta, \tilde{A})}_{4}.{\small III}$} & ~~~~$ds^{2}=\rho dx^{2}-2 dx dv+2 e^{x} dy du-\rho \eta^{2}e^{2x}du^{2},$ & \\& ~~~~$B=\kappa y e^{x} du\wedge dx+\tilde{A} e^{x}dv\wedge du$ & $\omega=0,~\lambda=1$  \\
	
				{$H^{(\kappa, \eta, \tilde{A})}_{4}.IV$} & ~~~~$ds^{2}=\frac{1}{1-\eta^{2}}\big[\rho  dx^{2}-2 dx dv-2\eta^{2}  y e^{x} dx du\big] +2 e^{x} dy du,$ & \\&
~~~~$B=(\kappa- \frac{\tilde{A}}{1-\eta^{2}}) ye^{x} du\wedge dx$ &  $\omega=0,~\lambda=1$\\
	
				{$H^{(\kappa, \eta, \tilde{A})}_{4}.V$} & ~~~~$ds^{2}=\rho  dx^{2}- 2 dx dv+ 2 e^{x} dy du - \frac{4 \eta^{2}}{1-\eta^{2}} e^{x} dx du,$ & \\&~~~~ $B=(\kappa + \tilde{A}) y e^{x} du\wedge dx$  & $\omega=-1,~\lambda=1$\\
	
				{$H^{(\kappa, \eta, \tilde{A})}_{4}.VI$} & ~~~~$ds^{2}=\rho  dx^{2}-2 dx dv+2 e^{x} dy du-\frac{2 \rho \eta^{2}}{1-\eta^{2}}   e^{x} dx du-\frac{ \rho \eta^{2}}{1-\eta^{2}}e^{2x}du^{2},$ & \\&~~~~ $B=(\kappa + \tilde{A}) y e^{x} du\wedge dx  +\tilde{A}  e^{x} dv \wedge du$ & $\omega=-1,~\lambda=1$\\
	
				{$H^{(\kappa, \eta, \tilde{A})}_{4}.VII$} &~~~~ $ds^{2}=\frac{\rho}{1+\eta^{2}} dx^{2}-2 dx dv +2 e^{x} dy du-\frac{\rho \eta^{2}}{1+\eta^{2}}   e^{2x}  du^{2},$&  \\& ~~~~$B= \kappa y e^{x} du\wedge dx + \tilde{A}e^{x}  dv\wedge du$ & $\omega=1,~\lambda=1$\\
	
				{$H^{(\kappa, \eta)}_{4}.VIII$} &~~~~ $ds^{2}= (1-\eta^{2})(\rho dx^{2}-2 dx dv)+2 e^{x} dy du+ 2 \eta^{2} y e^{x} dx du,$ & \\& ~~~~$B=\kappa y e^{x} du\wedge dx$ & $\omega=-1,~\lambda=1$\\
	
				{$H^{(\kappa, \eta, \tilde{A})}_{_{4,q}}.IX$} &~~~~ $ds^{2}=\frac{1-\eta^{2}}{1-\eta^{2}q^{4}}(\rho dx^{2}-2 dx dv)+2 e^{x} dy du+  \frac{2\eta^{2}(1-q^{4})}{1-\eta^{2}q^{4}} y e^{x} dx du,$ &  \\& ~~~~$B=\Big[\kappa-\frac{\tilde{A}  q^{2} (1-\eta^{2})}{1-\eta^{2}q^{4}}\Big] y e^{x} du\wedge dx$ & $\omega=-1,~\lambda=1$\\
	
				{$H^{(\kappa, \tilde{A})}_{4}.{\small X}$} & ~~~~$ds^{2}=\rho  dx^{2}-2 dx dv+2 e^{x} dy du,$ &\\& ~~~~$B=(\kappa - \tilde{A}) y e^{x} du\wedge dx$  &  $\omega=-1,~\lambda=1$\\  \hline \hline
	
		\end{tabular}}}
\end{center}
\vspace{-3mm}
{\scriptsize ~~~~~~~~~~~~~~~~~~$^\ast$Here we have ignored the total derivative terms that appeared in the $B$-fields part.}
\\\\
Finally, we find that the metric of background $H^{(\kappa, \eta, \tilde{A})}_{_{4,q}}.IX$ can be equal to \eqref{3.8} if
one applies the transformation
\begin{eqnarray}\label{3.34}
x^{'}=\frac{1-\eta^{2}}{1-\eta^{2} q^4} x,~~~~~y^{'}= y ~e^{\frac{\eta^{2} (1-q^4)}{1-\eta^{2} q^4}  x},~~~~u^{'} =u,~~~~v^{'}=v,
\end{eqnarray}
and also $\rho'=\rho (1-\eta^{2} q^4)/ (1 - \eta^2)$. Thus, we showed that, in some cases of the deformed backgrounds,
the $H_{4}$ WZW model metric is, under arbitrary YB deformations, invariant up to antisymmetric $B$-fields.
\\\\
{\bf The backgrounds $H^{(\kappa, \eta, \tilde{A})}_{4}.III, H^{(\kappa, \eta, \tilde{A})}_{4}.VI$ and $H^{(\kappa, \eta, \tilde{A})}_{4}.VII$}.
In order to clarify the structure of the metrics of $H^{(\kappa, \eta, \tilde{A})}_{4}.III, H^{(\kappa, \eta, \tilde{A})}_{4}.VI$ and $H^{(\kappa, \eta, \tilde{A})}_{4}.VII$
one may find isometry group of the metrics, where the generators of the corresponding Lie algebra
can be expressed in terms of the Killing vectors. One immediately find that
the metrics of these backgrounds admit a six-dimensional Lie algebra of Killing vectors, which it cannot evidently be isomorphic to those of \eqref{3.8.2}. Accordingly, these backgrounds cannot be turned into the $H_{4}$ WZW model.

%%%%%%%%%%%%%%%%%%%%%%%%%%%%%%%%%%%%%%%%%%%%%%%%%%%%%%%%%%%%%%%%%%%%%%%%%%%%%%%%%%%%%%%%%%%%%
\section{\label{Sec.IV}Conformal invariance of the backgrounds up to two-loop}

In the $\sigma$-model context, the conformal invariance conditions of the $\sigma$-model are
provided by the vanishing of the beta-function equations \cite{callan1985strings}.
The study of the conformal invariance has led to the covering of string theory,
since one- and two-loop domains in string theory correspond to formulating on worldsheets of nontrivial topology.
It is well known that the conditions for conformal invariance can be interpreted as effective field equations for $G_{\mu \nu}$,  $B_{\mu \nu}$ and dilaton field $\Phi$ of the string effective action \cite{callan1985strings}.
The dilaton field is only one more massless degree of freedom of the bosonic string theory.
This gives a contribution to the action \eqref{3.7} in the form of $ \frac{1}{8\pi} \int d\tau  d\sigma  R^{^{(2)}} \Phi(x^\mu)$
in which $R^{^{(2)}}$ is the curvature scalar on the string worldsheet. This term breaks Weyl invariance on a classical level as do the one-loop corrections to ${G}$ and  ${B}$.
Below, we shall solve the one- and two-loop beta-function equations for all YB deformed backgrounds of Table 1 to obtain
the corresponding dilaton fields\footnote{Notice that there is a one-to-one correspondence between the r-matrices $ r_{_{I}},  r_{_{II}},  r_{_{IV}}$ as solutions of the CYBE and two-dimensional Abelian subalgebra. These solutions satisfy the unimodularity condition of \cite{supergeometry,Two-loop} while
for the case of $ r_{_{III}}$, two-dimensional subalgebra is non-Abelian; accordingly, the unimodularity condition is not satisfied.
Anyway we still have a solution for which the conformal invariance condition is satisfied at one-loop level, as well as two-loop.
Also, one can check the two-loop conformal invariance conditions for YB deformed backgrounds constructed from the
 matrices $ r_{_{V}},...,  r_{_{X}}$. Here we do not have the condition of \cite{supergeometry}, because of the existence of a WZW term.}.

\subsection{\label{subSec.IV.1}Conditions for one-loop solution}

The conditions for conformal invariance to hold in the $\sigma$-model in the lowest nontrivial approximation
are the vanishing of the one-loop beta-function.
The equations for the vanishing of the one-loop beta-function are given by \cite{callan1985strings}
\begin{eqnarray}
0 &=&R_{{\mu \nu}}-(H^2)_{\mu \nu}+{\nabla}_\mu
{\nabla}_\nu \Phi,\nonumber\\
0 &=&-{\nabla}^\lambda H_{{\lambda \mu \nu}} + H_{{\mu \nu}}^{~\;\lambda}  ~{\nabla}_\lambda\Phi,\nonumber\\
0 &=&2\Lambda + {\nabla}^2 \Phi' - ({\nabla} \Phi')^2+\frac{2}{3}  H^{{2}},\label{4.1}
\end{eqnarray}
where  $R_{{\mu \nu}}$ is the Ricci tensor of the metric $G_{\mu \nu}$,  $H_{_{\mu \nu \rho}}$ defined by
\begin{eqnarray}\label{4.2}
H_{{\mu \nu \rho}} = \frac{1}{2} \big(\partial_{\mu} B_{\nu \rho}+\partial_{\nu} B_{\rho \mu}+\partial_{\rho} B_{\mu \nu}\big),
\end{eqnarray}
is the torsion of the antisymmetric $B$-field, and $\Lambda$ is a cosmological constant which vanishes for critical strings.
We have also used the conventional notations
 $(H^2)_{\mu \nu} = H_{{\mu \rho \sigma }} H^{{\rho \sigma}}_{~~\nu}$ and $H^2 = H_{{\mu \nu \rho}} H^{{\mu \nu \rho}} $.
We now solve the field equations \eqref{4.1} for all YB deformed backgrounds of Table 1. In this way, we find the dilaton fields that
guarantee the conformal invariance of the backgrounds at one-loop level. In all cases, the cosmological constant vanishes.
In order to get more clarity, the results obtained for the dilaton fields are summarized in Table 2.

\vspace{1cm}

		\begin{center}	
			{\scriptsize { {{\bf Table 2.} The dilaton fields making the $H_4$ deformed backgrounds conformal up to one-loop order$^\ast$}} }\\
			\renewcommand{\arraystretch}{1.5}{{\footnotesize 	
					\begin{tabular}{p{1.5cm}ll} \hline \hline
						
						Background symbol &~~~~~~~~~~~~~~~Dilaton fields & ~~~~~~~~~~~~~~~~~~Comments\\ \hline
						
						$H^{(\kappa)}_{4}.I$ & $~~~~~~~~~~~~~~~\frac{1}{4}(1-\kappa^{2})x^{2} + c_{1} x +c_{2}$ &  \\
						
						$H^{(\kappa, \eta)}_{4}.II$ & $~~~~~~~~~~~~~~~\frac{1}{4}(1-\kappa^{2})x^{2} + c_{1} x +c_{2} $ &  \\
						
						$H^{(\kappa, \eta, \tilde{A})}_{4}.III$ & $~~~~~~~~~~~~~~~\frac{1}{4}\big(1-\kappa^{2}\big)x^{2} + c_{1} x +c_{2}$ & $~~~~~~~~~~~~~~~~~~{\tilde A}=0$  \\
						
						$H^{(\kappa, \eta, \tilde{A})}_{4}.IV$ & $~~~~~~~~~~~~~~~\frac{1}{4}\Big[\frac{1}{(\eta^{2}-1)^{2}}-\big(\kappa+\frac{\tilde{A}}{\eta^{2}-1}\big)^{2}\Big]x^{2} + c_{1} x +c_{2}$  \\
						
						$H^{(\kappa, \eta, \tilde{A})}_{4}.V$ & $~~~~~~~~~~~~~~~\frac{1}{4}\Big[1-(\kappa +\tilde{A})^2\Big] x^{2} + c_{1} x +c_{2}$  \\
						
						$H^{(\kappa, \eta, \tilde{A})}_{4}.VI$ & $~~~~~~~~~~~~~~~\frac{1}{4}\big(1-\kappa^{2}\big)x^{2} + c_{1} x +c_{2}$ & $~~~~~~~~~~~~~~~~~~{\tilde A}=0$  \\
						
						$H^{(\kappa, \eta, \tilde{A})}_{4}.VII$ & $~~~~~~~~~~~~~~~\frac{1}{4}\big(1-\kappa^{2}\big)x^{2} + c_{1} x +c_{2}$ & $~~~~~~~~~~~~~~~~~~{\tilde A}=0$\\
						
						$H^{(\kappa, \eta)}_{4}.VIII$ & $~~~~~~~~~~~~~~~\frac{1}{4}\big[(1-\eta^{2})^{2}-\kappa^{2}\big]x^{2} + c_{1} x +c_{2}$ \\
						
						$H^{(\kappa, \eta, \tilde{A})}_{_{4,q}}.IX$ & $~~~~~~~~~~~~~~~\frac{1}{4}\Big[(\frac{1-\eta^{2}}{1-\eta^{2} q^{4}})^2
						-\big(\kappa-\frac{\tilde{A}q^2(1-\eta^{2})}{1-\eta^{2} q^{4}}\big)^{2}\Big]x^{2} + c_{1} x +c_{2}$   \\
						
						$H^{(\kappa, \tilde{A})}_{4}.X$ & $~~~~~~~~~~~~~~~\frac{1}{4}\big[1-(\tilde A-\kappa)^{2}\big] x^{2} + c_{1} x +c_{2}$  \\ \hline \hline	 
			\end{tabular}}}
		\end{center}
\vspace{-3mm}
{\scriptsize ~~~~~~~~~~~~~~~~$^\ast$ Here $c_1$ and $c_2$ are some arbitrary constants.}\\\\

\subsection{\label{subSec.IV.2}Conditions for two-loop solution}

In order for the fields $(G, B, \Phi)$ to provide a consistent string background at low-energy
up to two-loop order, they must satisfy the following equations \cite{hull1988string,metsaev1987two}
\begin{eqnarray}\label{4.3}
0 &=&R_{{\mu \nu}}-(H^2)_{\mu \nu}+{\nabla}_\mu
{\nabla}_\nu \Phi +\frac{1}{2} \alpha' \Big[R_{{\mu \rho \sigma \lambda}} R_{\nu}^{{~\rho\sigma \lambda}}
+2 R_{{\mu \rho\sigma \nu }} (H^2)^{\rho\sigma}\nonumber\\
~~&~~~&+2 R_{{\rho\sigma \lambda(\mu }}H_{\nu)}^{~\lambda \delta} H^{\rho\sigma}_{~~_{\delta}} +\frac{1}{3} ({\nabla}_\mu H_{\rho\sigma\lambda})
({\nabla}_\nu H^{\rho\sigma\lambda}) -({\nabla}_\lambda H_{\rho\sigma \mu})
({\nabla}^\lambda H^{\rho \sigma}_{~~\nu})\nonumber\\
~~&~~~&+2 H_{{\mu \rho \sigma}} H_{{\nu \lambda \delta}} H^{{\eta \delta \sigma}} H_{\eta}^{~~\lambda \rho}
+2 H_{{\mu \sigma \lambda}}  H_{{\nu \rho}} ^{~~\lambda}  (H^2)^{\rho\sigma} \Big]+
{\cal O}(\alpha'^2),\nonumber\\
0 &=&{\nabla}^\lambda H_{{\lambda \mu \nu}} -  ({\nabla}^\lambda\Phi')  H_{{\mu \nu \lambda}}
+ \alpha' \Big[{\nabla}^\lambda H^{\rho \sigma}_{~~[\mu}R_{_{\nu]} \lambda \rho \sigma} - ({\nabla}_\lambda H_{\rho\mu\nu}) (H^2)^{\lambda\rho}\nonumber\\
~~&~~~&- 2 ({\nabla}^\lambda H^{\rho \sigma}_{~~[\mu})H_{_{\nu]} \rho \delta } H_{\lambda \sigma}^{~\;\delta}\Big]
+{\cal O}(\alpha'^2),\nonumber\\
0 &=&2\Lambda + {\nabla}^2 \Phi' - ({\nabla} \Phi')^2+\frac{2}{3}  H^{{2}}
-\alpha' \Big[\frac{1}{4} R_{{\mu \rho \sigma \lambda}} R^{{\mu \rho \sigma \lambda}}\nonumber\\
&~~~-&\frac{1}{3} ({\nabla}_\lambda H_{\mu \nu \rho })
 ({\nabla}^\lambda H^{\mu \nu \rho })
-\frac{1}{2} H^{\mu\nu}_{~~\lambda} H^{\rho \sigma \lambda} R_{\mu \nu \rho \sigma } -R_{\mu \nu} (H^2)^{\mu \nu} +\frac{3}{2} (H^2)_{\mu \nu} (H^2)^{\mu \nu}\nonumber\\
&~~~+& \frac{5}{6} H_{\mu \nu \rho } H^{\mu}_{~~\sigma \lambda} H^{\nu \sigma}_{~~\delta} H^{\rho \lambda \delta}\Big] +{\cal O}(\alpha'^2),
\end{eqnarray}
where ${R}_{{\mu\nu\rho\sigma}}$ is the
Riemann tensor field of the metric $G_{{\mu\nu}}$, $(H^2)^{\mu \nu} = H^{\mu \rho \sigma } H_{\rho \sigma}^{~~\nu}$, and in second equation of \eqref{4.3}
$\Phi' = \Phi + \alpha' q H^2$ for some coefficient $q$ \cite{hull1988string}.
We note that round brackets denote the symmetric part on the indicated indices whereas square brackets
denote the antisymmetric part.
Using the above equations we check the conformal invariance conditions of
the backgrounds of Table 1 up to two-loop order.
In fact, we introduce some new solutions for two-loop beta-function equations of the $\sigma$-model with a non-vanishing field strength
$H$ and the dilaton field in the absence of a cosmological constant $\Lambda$.
The field equations \eqref{4.3} are satisfied for all backgrounds of Table 1 with the same dilaton fields given in Table 2.

%%%%%%%%%%%%%%%%%%%%%%%%%%%%%%%%%%%%%%%%%%%%%%%%%%%%%%%%%%%%%%%%%%%%%%%%%%%%%%%%%%%%%%%%%%%%%%%%%%%%%

\section{\label{Sec.V}Summary and concluding remarks}

Using automorphism group of the $ h_{4} $ Lie algebra
we have classified all corresponding classical r-matrices as the solutions of (m)CYBE.
Then, we obtained all YB deformed WZW models based on the $H_{4}$ Lie group.
We have, in some cases, shown that the metric of the $H_{4}$ WZW model is invariant under possible YB deformations while
the antisymmetric $B$-fields are changed.
We have also shown that all new integrable backgrounds of YB deformed $H_{4}$
WZW model are conformally invariant up to two-loop in the absence of a cosmological constant $\Lambda$.
In this respect, we have derived the general form of the dilaton fields satisfying the
vanishing beta-function equations. In fact, the YB deformed backgrounds
that are conformal at one-loop remain conformal at two-loop with the same dilaton fields.
{\it Most importantly, it has been shown that the $H_{4}$ WZW model is a
conformal theory within the class of the YB deformations preserving the conformal
invariance up to two-loop order.}
It is also straightforward to determine the dilaton in the YB deformed Nappi-Witten model \cite{kyono2016yang}
by following our present analysis and method. As it has been indicated in Appendix B, we have classified all
nonequivalent r-matrices of the Nappi-Witten Lie algebra in order to study
the corresponding YB deformation of WZW model.

As a future direction, it would be interesting to generalize the YB deformation formulation of WZW model from Lie groups to Lie supergroups.
As we know already, in order to construct the YB deformations of WZW model on a Lie group $G$
one needs the r-matrices of Lie algebra $\G$ of $G$. Fortunately, the classical r-matrices related to some of the
Lie superalgebras are available \cite{{J.z},{J},{ER6},{ER14}} (see also \cite{ER3}). One can use these to construct new backgrounds of YB deformed
WZW models. We hope that in future it will be possible to
find YB deformed WZW models even for physically interesting backgrounds.
The generalization of YB deformation of WZW model to Lie supergroups is currently under investigation.

\subsection*{Acknowledgements}
The authors would like to thank the anonymous referee for invaluable comments and criticisms.
This work has been supported by the research vice
chancellor of Azarbaijan Shahid Madani University under research fund No. 97/231.
%%%%%%%%%%%%%%%%%%%%%%%%%%%%%%%%%%%%%%%%%%%%%%%%%%%%%%%%%%%%%%%%%%%%%%%%%%%%%%%%%%%%%%%%%%%%%%%%%%%%%%%
\appendix

\section{\label{app.A}Some computational results related to YB deformations of the $H_{4}$ WZW model}

In this appendix, we tabulate the nonzero components of tensors $ H_{\mu\nu\rho}, (H^{2})_{\mu\nu}, R_{_{\mu \nu}}$ and Riemann tensor field
related to the backgrounds of YB deformed $H_4$ WZW model representing in Table 1.
We note that for all backgrounds one quickly finds that $R=H^{2}=0$; moreover,
the only nonzero component of $R_{_{\mu \nu}}$ is $R_{_{xx}}$ which is indicated for all backgrounds in Table 3.
\begin{center}
	\scriptsize {{{\bf Table 3.} ~The nonzero components of tensors $R_{_{\mu\nu}}, R_{_{\mu \nu \rho \sigma}}, H_{_{\mu\nu\rho}}$ and $(H^2)_{_{\mu\nu}}$
			related to the backgrounds represented in Table 1}}\\
	{\scriptsize
		\renewcommand{\arraystretch}{1.5}{
			\begin{tabular}{p{1.5cm}llll} \hline \hline
				Background symbol & ~~$R_{_{xx}}$ & $R_{_{\mu \nu \rho \sigma}}$ & $H_{_{\mu\nu\rho}}$ & $(H^2)_{_{\mu\nu}}$ \\ \hline

				$H^{(\kappa}_{4}.I$  & ~~ - $\frac{1}{2}$   &   $R_{_{xyxu}}=\frac{-e^x}{4}$    &    $H_{_{xyu}} = \frac{\kappa e^x}{2}$    &   $(H^2)_{_{xx}} = \frac{-\kappa^2}{2}$ \\\\

				$H^{(\kappa, \eta)}_{4}.II$  &  ~~ -$\frac{1}{2}$   &   $R_{_{xyxu}}=\frac{- e^x}{4}$  & $H_{_{xyu}} = \frac{\kappa e^x}{2}$
 & $(H^2)_{_{xx}} = -\frac{\kappa^2}{2 }$\\\\

				$H^{(\kappa, \eta, \tilde{A})}_{4}.III$  & ~~  -$\frac{1}{2}$   &   $R_{_{xyxu}}=\frac{-e^x}{4},$  & $H_{_{xyu}} = \frac{\kappa e^x}{2}$  & $(H^2)_{_{xx}} = \frac{-\kappa^2}{2}$
				\\
				& ~~     &   $R_{_{xuxu}}=\frac{5 \rho \eta^2  e^{2x}}{4}$  & $H_{_{xuv}} = \frac{- \tilde A e^x}{2}$  & $(H^2)_{_{xu}} = \frac{-\kappa \tilde A}{2} e^x$
				\\& ~~     &     &  & $(H^2)_{_{uu}} = \frac{- {\tilde A}^2}{2} e^{2x}$\\\\

				$H^{(\kappa, \eta, \tilde{A})}_{4}.IV$  & ~~  -$\frac{1}{2(1-\eta^2)^2}$   &   $R_{_{xyxu}}=\frac{-e^x}{4(1-\eta^2)^2}$  & $H_{_{xyu}} = \frac{1}{2} (\kappa-\frac{\tilde A}{1-\eta^2}) e^x$  & $(H^2)_{_{xx}} = \frac{-1}{2} \big(\kappa-\frac{\tilde A}{1-\eta^2}\big)^2$\\\\

				$H^{(\kappa, \eta, \tilde{A})}_{4}.V$  & ~~  -$\frac{1}{2}$   &   $R_{_{xyxu}}=-\frac{e^x}{4}$  & $H_{_{xyu}} = \frac{1}{2} (\kappa+\tilde{A})e^{x}$  & $(H^2)_{_{xx}} =\frac{-(\kappa+\tilde{A})^{2}}{2}$\\\\

				$H^{(\kappa, \eta, \tilde{A})}_{4}.VI$  & ~~  -$\frac{1}{2}$   &   $R_{_{xyxu}}=\frac{-e^x}{4},$  & $H_{_{xyu}} = \frac{1}{2} (\kappa+\tilde{A})e^{x}$  & $(H^2)_{_{xx}} = \frac{-(\kappa+\tilde{A})^{2}}{2}$
				\\
				& ~~     &   $R_{_{xuxu}}=\frac{5\rho\eta^{2}e^{2x}}{4(1-\eta^{2})}$  & $H_{_{xuv}} = \frac{-\tilde{A}e^{x}}{2}$  & $(H^2)_{_{xu}} = \frac{-(\kappa+\tilde{A})\tilde{A}e^{x}}{2}$
				\\
				& ~~     &     &  & $(H^2)_{_{uu}} = \frac{- {\tilde A}^2}{2} e^{2x}$\\\\

				$H^{(\kappa, \eta, \tilde{A})}_{4}.VII$  & ~~  -$\frac{1}{2}$   &   $R_{_{xyxu}}=\frac{-e^x}{4},$  & $H_{_{xyu}} = \frac{\kappa e^{x}}{2}$  & $(H^2)_{_{xx}} = \frac{-\kappa^{2}}{2}$
				\\
				& ~~     &   $R_{_{xuxu}}=\frac{5\rho\eta^{2}e^{2x}}{4(1+\eta^{2})}$  & $H_{_{xuv}} = \frac{-\tilde{A}e^{x}}{2}$  & $(H^2)_{_{xu}} = \frac{-\kappa\tilde{A}e^{x}}{2}$
				\\
				& ~~     &     &  & $(H^2)_{_{uu}} = \frac{- {\tilde A}^2}{2} e^{2x}$\\\\

				$H^{(\kappa, \eta)}_{4}.VIII$  & ~~  -$\frac{(1-\eta^{2})^{2}}{2}$   &   $R_{_{xyxu}}=-\frac{e^{x}(1-\eta^{2})^{2}}{4}$    &    $H_{_{xyu}} = \frac{\kappa e^{x}}{2}$    &   $(H^2)_{_{xx}} = -\frac{\kappa^{2}}{2}$ \\\\

				$H^{(\kappa, \eta, \tilde{A})}_{_{4,q}}.IX$  & ~~  -$\frac{1}{2}\big(\frac{1-\eta^{2}}{1-\eta^{2}q^{4}}\big)^2$   &   $R_{_{xyxu}}=-\frac{1}{4}\big(\frac{1-\eta^{2}}{1-\eta^{2}q^{4}}\big)^2 e^x$    &    $H_{_{xyu}} = \frac{e^x}{2} \big(\kappa-\frac{\tilde{A}q^2(1-\eta^{2})}{1-\eta^{2}q^{4}}\big) $    &   $(H^2)_{_{xx}} = -\frac{1}{2} {\big(\kappa-\frac{\tilde{A}q^{2}(1-\eta^{2})}{1-\eta^{2}q^{4}}\big)^{2}}$ \\\\

				$H^{(\kappa, \tilde{A})}_{4}.X$  & ~~ - $\frac{1}{2}$   &   $R_{_{xyxu}}=\frac{-e^x}{4}$    &    $H_{_{xyu}} = \frac{1}{2}(\kappa -\tilde A)e^x$    &
				$(H^2)_{_{xx}} = \frac{-1}{2}(\kappa -\tilde A)^2$
				\\    \hline \hline
				
	\end{tabular}}}
\end{center}

\section{\label{app.B}More on YB deformations of the Nappi-Witten WZW model}

\subsection{\label{app.B.1}Nonequivalent r-matrices}

In this appendix, using automorphism group of the Nappi-Witten Lie algebra \cite{rezaei2010complex,christodoulakis2002automorphisms}
we find all nonequivalent r-matrices as solutions of the (m)CYBE \eqref{3.18}.
We then find all YB deformations of WZW model on the Nappi-Witten Lie group.
Before proceeding to get nonequivalent r-matrices, let us introduce the Nappi-Witten Lie algebra.
It is spanned by the set of generators $T_{_i} =(P_{_1}, P_{_2}, J, T)$ which
fulfill the following nonzero commutation rules \cite{nappi1993wess}:
\begin{eqnarray}\label{B.1}
[J , P_{_1}]= P_{_2},~~~~~~[J , P_{_2}]=- P_{_1},~~~~~~[P_{_1} , P_{_2}]= T.
\end{eqnarray}
This algebra is a central extension of the 2D Poincar\'{e} algebra to which it reduces if one sets $T= 0$.
Using \eqref{B.1} together with \eqref{2.7},
one obtains the non-degenerate ad-invariant bilinear form $\Omega_{ij}$ on the Nappi-Witten Lie algebra, giving
\begin{eqnarray}\label{B.2}
\Omega_{ij} = \begin{pmatrix}
					1 &~0~&~0~& 0\;\\
					0 & 1 & 0 & 0\\
					0 & 0 & b & 1\\
					0 & 0 & 1 & 0
				\end{pmatrix},
\end{eqnarray}
where $b$ is a real constant.
In order to calculate the  left-invariant one-forms
${L}_{\alpha}$  we parameterize the Nappi-Witten group with coordinates
$x^{\mu} =(a_{_1}, a_{_2}, u, v)$ so that its elements can be written as
\begin{eqnarray}\label{B.3}
g=\exp\bigl(a_{_1} P_{_1}+a_{_2} P_{_2}\bigr)\exp\bigl(u\,J+v\,T\bigr).
\end{eqnarray}
We then obtain
\begin{eqnarray}\label{B.4}
L_{\alpha}=g^{-1}~\partial_\alpha g &=& \big(\cos u\, \partial_\alpha a_1+\sin u\, \partial_\alpha a_2\big)\,P_{_1}
			+\big(\cos u\, \partial_\alpha a_2-\sin u\, \partial_\alpha a_1\big)\,P_{_2}\nonumber\\
			&&+\partial_\alpha u\, J+\big(\partial_\alpha v + \frac{1}{2}\,a_2\, \partial_\alpha a_1-\frac{1}{2}\,a_1\,\partial_\alpha a_2\big)\,T.
\end{eqnarray}
Using the above results together with the general form of the WZW model action \eqref{3.2},
the spacetime metric and antisymmetric $B$-field are, respectively,  found to be
\begin{eqnarray}\label{B.5}
ds^2 &=& 2dudv+b\,du^2 +d{a_{_1}}^2+d{a_{_2}}^2-a_{_1}\,da_{_2} du+a_{_2}\,da_{_1} du,\nonumber\\
B &=& u~da_{_1} \wedge da_{_2}.	
\end{eqnarray}
According to \eqref{2.6}, \eqref{2.8} and \eqref{3.10} to construct the YB deformation of WZW model on the Nappi-Witten group
we need to find the corresponding nonequivalent r-matrices. Using relations \eqref{B.1} and \eqref{B.2}
one may obtain the general solution of the (m)CYBE \eqref{3.18} as follows \cite{kyono2016yang}
\begin{eqnarray}\label{B.6}
r^{ij}=\left(\begin{tabular}{cccc}
                 0 & $\pm \sqrt{\omega}$ & 0 & $m_3$ \\
                 $\mp\sqrt{\omega}$ & 0  & 0 & $m_5$ \\
                 0& 0 & 0 & $m_6$ \\
                 -$m_3$ & -$m_5$ & -$m_6$ & 0 \\
                 \end{tabular}\right),
\end{eqnarray}
for some real constants $m_3, m_5, m_6$.
As was mentioned earlier, to obtain the nonequivalent r-matrices
one must use the automorphism group of the Nappi-Witten algebra.
Using \eqref{B.1} and \eqref{3.17} the automorphism groups of the Nappi-Witten algebra
are expressed as matrices in the following form \cite{{christodoulakis2002automorphisms},{rezaei2010complex}}
\begin{eqnarray}\label{B.7}
A_1=\left(\begin{tabular}{cccc}
                 a & b & 0 & -ac -bd \\
                 -b & a & 0 & -ad+bc \\
                 c & d & 1 & e \\
                 0 & 0 & 0 & $a^2+b^2$
                  \\
                 \end{tabular} \right),~~~~A_2=\left( \begin{tabular}{cccc}
                 a & b & 0 & ac +bd \\
                 b & -a & 0 & -ad+bc \\
                 c & d & -1 & e \\
                 0 & 0 & 0 & -$(a^2+b^2)$
                  \\
                 \end{tabular} \right),~~a^2+b^2 \neq0,
\end{eqnarray}
where $a, b, c, d$ and $e$ are some arbitrary constants. Ultimately, by employing formula \eqref{3.19} of Proposition 3.1,
the r-matrices for the Nappi-Witten algebra are split into the following six nonequivalent families
\begin{eqnarray*}
r'_{_1}=\left( \begin{tabular}{cccc}
                 0 & 0 & 0 & 1 \\
                 0 & 0 & 0 & 0 \\
                 0 & 0 & 0 & 0 \\
                 -1 & 0 & 0 & 0 \\
                 \end{tabular} \right),~~~r'_{_2}=\left( \begin{tabular}{cccc}
                 0 & 0 & 0 & 0 \\
                 0 & 0 & 0 & 0 \\
                 0 & 0 & 0 & 1 \\
                 0 & 0 &-1 & 0 \\
                 \end{tabular} \right),~~~r'_{_3}=\left( \begin{tabular}{cccc}
                 0 & 0 & 0 & 0 \\
                 0 & 0 & 0 & 0 \\
                 0 & 0 & 0 & -1 \\
                 0 & 0 &1 & 0 \\
                 \end{tabular} \right),
\end{eqnarray*}
\begin{eqnarray}\label{B.8}
r'_{_4}=\left( \begin{tabular}{cccc}
                 0 & 1 & 0 & 0 \\
                 -1& 0 & 0 & 0 \\
                 0 & 0 & 0 & 0 \\
                 0 & 0 & 0 & 0 \\
                 \end{tabular} \right),~~~r'_{_5}=\left( \begin{tabular}{cccc}
                 0 & 1 & 0 & 0 \\
                 -1& 0 & 0 & 0 \\
                 0 & 0 & 0 & $p^2$ \\
                 0 & 0 & -$p^2$ & 0 \\
                 \end{tabular} \right),~~~r'_{_6}=\left( \begin{tabular}{cccc}
                 0 & -1 & 0 & 0 \\
                 1& 0 & 0 & 0 \\
                 0 & 0 & 0 & -$p^2$ \\
                 0 & 0 & $p^2$ & 0 \\
                 \end{tabular} \right),
\end{eqnarray}
where $p$ is a nonzero constant.

\subsection{\label{app.B.2}Backgrounds for YB deformations of the Nappi-Witten WZW model}

Hence it is straightforward to study YB deformations of the Nappi-Witten WZW model.
Similar to the YB deformations of the $H_4$ WZW model in Sec. \ref{Sec.III},
we use formulas \eqref{3.12}, \eqref{3.13} and \eqref{B.2}
to obtain all linear $R$-operators corresponding to the nonequivalent r-matrices of the Nappi-Witten algebra.
Then, by using \eqref{2.6} together with \eqref{2.8} one obtains all YB deformed backgrounds of the Nappi-Witten WZW model.
The deformed backgrounds including metric and $B$-field are summarized in Table 4.
\begin{center}
	\scriptsize {{{\bf Table 4.}~ YB deformed backgrounds of the Nappi-Witten WZW model$^\ast$}}
	{\scriptsize
		\renewcommand{\arraystretch}{1.5}{
			\begin{tabular}{p{1.5cm}ll} \hline \hline
				Nonequivalent r-matrices & ~~~~Backgrounds & Comments\\ \hline
				{$r'_{_1}$} & ~~~~$ds^2=da_1^2+da_2^2+(b-\eta^2 )du^2+2du dv+a_2 da_1 du-a_1 da_2du,$ \\& ~~~~$B=\kappa\,u\, da_1\wedge da_2$  & $\omega=0$\\
				
				{$r'_{_2}$} & ~~~~$ds^2=da_1^2+da_2^2+\frac{1}{1-\eta^{2}} \big [bdu^2+2 du\, dv+ a_2 da_1 du-a_1 da_2du\big],$ \\& ~~~~$B = \kappa\,u\,da_1\wedge da_2+\frac{\tilde{A}}{2(1-\eta^2)}
				\big[a_2\,du\wedge da_1+a_1\,da_2\wedge du\big]$ & $\omega=0$ \\
				
				{$r'_{_3}$} & ~~~~$ds^2=da_1^2+da_2^2+\frac{1}{1-\eta^{2}} \big [bdu^2+2 du\, dv+ a_2 da_1 du-a_1 da_2du\big],$ \\& ~~~~$B = \kappa\,u\,da_1\wedge da_2-\frac{\tilde{A}}{2(1-\eta^2)}
				\big[a_2\,du\wedge da_1+a_1\,da_2\wedge du\big]$ & $\omega=0$\\
				{$r'_{_4}$} &~~~~ $
				ds^2= da_1^2+da_2^2+(1+\eta^2) \big[b du^2+2 du\,dv
				+a_2 da_1 du-a_1 da_2du\big],
				$ \\&~~~~ $B=\kappa\,u\,da_1\wedge da_2$& $\omega=1$ \\
				
				{$ r'_{_5}$} & ~~~~$
				ds^2= da_1^2+da_2^2+\frac{1+\eta^2}{1-\eta^{2} p^4} \big[b du^2+2 du\,dv
				+a_2 da_1 du-a_1 da_2du\big],
				$ \\&~~~~ $B=\kappa\,u\,da_1\wedge da_2+\frac{\tilde{A} p^2(1+\eta^2)}{2(1-\eta^{2} p^4)}
				\big[a_2\,du\wedge da_1+a_1\,da_2\wedge du\big]$ & $\omega=1$ \\
				{$r'_{_6}$} &~~~~ $
				ds^2= da_1^2+da_2^2+\frac{1+\eta^2}{1-\eta^{2} p^4} \big[b du^2+2 du\,dv
				+a_2 da_1 du-a_1 da_2du\big],
				$ \\& ~~~~$B=\kappa\,u\,da_1\wedge da_2-\frac{\tilde{A} p^2(1+\eta^2)}{2(1-\eta^{2} p^4)}
				\big[a_2\,du\wedge da_1+a_1\,da_2\wedge du\big]$& $\omega=1$ \\  \hline \hline		
				
	\end{tabular}}}
\end{center}
\vspace{-3mm}
{\scriptsize ~~~~~~~~~~~~~~~~~~~~~~~$^\ast$Here we have ignored the total derivative terms that appeared in the $B$-fields part.}

\subsection{\label{app.B.3}Conformal invariance of the backgrounds up to one- and two-loop orders}

In order to guarantee the conformal invariance of the YB deformed backgrounds of
the Nappi-Witten WZW model of Table 4, at least at the one-loop level,
one must show that they satisfy the vanishing beta-function equations \eqref{4.1}.

\begin{center}	
	{\scriptsize { {{\bf Table 5.} The dilaton fields making the Nappi-Witten deformed backgrounds conformal up to the one- and two-loop orders}} }\\
	\renewcommand{\arraystretch}{1.5}{{\footnotesize 	
			\begin{tabular}{p{1.5cm}l} \hline \hline
				
				Nonequivalent r-matrices & ~~~~~Dilaton fields \\ \hline
				
				$r'_{_1}$ & $~~~~~\frac{1}{4}(\kappa^{2}-1)u^{2} + c_{1} u +c_{2} $    \\
				
				$r'_{_2}$  & $~~~~~\frac{1}{4}\big[\kappa^{2}-\frac{2 \kappa \tilde{A}}{\eta^{2}-1}-\frac{(1-\tilde{A}^{2})}{(\eta^{2}-1)^{2}}\big]u^{2} + c_{1} u +c_{2}$ \\
				
				$r'_{_3}$ & $~~~~~\frac{1}{4}\big[\kappa^{2}+\frac{2 \kappa \tilde{A}}{\eta^{2}-1}-\frac{(1-\tilde{A}^{2})}{(\eta^{2}-1)^{2}}\big]u^{2} + c_{1} u +c_{2}$   \\
				
				$r'_{_4}$  & $~~~~~\frac{1}{4}\big[\kappa^{2}-(1+\eta^{2})^{2}\big]u^{2} + c_{1} u +c_{2}$  \\
				
				$r'_{_5}$  & $~~~~~\frac{1}{4(1-\eta^{2} p^4)^{2}}\Big[\kappa^{2}(1-\eta^{2} p^4)^{2}+2 \kappa  \tilde{A} p^2(1+\eta^2)(1-\eta^{2} p^4)-(1- \tilde{A}^{2} p^{4})(1+\eta^2)^{2}\Big]u^{2} + c_{1} u +c_{2}$     \\
				$r'_{_6}$ & $~~~~~\frac{1}{4(1-\eta^{2} p^4)^{2}}\Big[\kappa^{2}(1-\eta^{2} p^4)^{2}-2 \kappa  \tilde{A} p^2(1+\eta^2)(1-\eta^{2} p^4)-(1- \tilde{A}^{2} p^{4})(1+\eta^2)^{2}\Big]u^{2} + c_{1} u +c_{2}$     \\ \hline \hline	
	\end{tabular}}}
\end{center}
From solving Eqs. \eqref{4.1} we find the general form
of the dilaton fields that make the YB deformed backgrounds conformal up to the one-loop order.
The results obtained for dilaton fields are represented in Table 5.
It would also be interesting to consider the conformal invariance of the Nappi-Witten deformed backgrounds up to the two-loop order.
To this end, we solve the field equations \eqref{4.3} and show the YB deformed backgrounds
 that are conformal at one-loop remain conformal at two-loop with the same dilaton fields given in Table 5. In this way,
the cosmological constant vanishes.

%%%%%%%%%%%%%%%%%%%%%%%%%%%%%%%%%%%%%%%%%%%%%%%%%%%%%%%%%%%%%%%%%%%%%%%%%%%%%%%
%%%%%%%%%%%%%%%%%%%%%%%%%%%%%%%%%%%%%%%%%%%%%%%%%%%%%%%%%%%%%%%%%%%%%%%%%%%%%

\end{document}